\documentclass[twocolumn,showpacs,preprintnumbers,amsmath,amssymb]{revtex4}

\usepackage{epsfig,psfrag}
\usepackage{dcolumn}
\usepackage{bm}
\usepackage{graphicx}
\usepackage{color}

\begin{document}
\title{ From Luttinger liquid to Altshuler-Aronov anomaly in multi-channel 
quantum wires  }
\author{Christophe Mora,$^1$ Reinhold Egger,$^1$ and Alexander Altland$^2$}
\affiliation{
${}^1$~Institut f\"ur Theoretische Physik,
Heinrich-Heine Universit\"at, D-40225 D\"usseldorf\\
${}^2$~Institut f\"ur Theoretische Physik,
Universit\"at zu K\"oln, Z\"ulpicher Str. 77, D-50937 K\"oln
}
\date{\today}
\begin{abstract}
  A crossover theory connecting Altshuler-Aronov electron-electron
  interaction corrections and Luttinger liquid behavior in quasi-1D
  disordered conductors has been formulated. Based on an interacting
  non-linear sigma model, we compute the tunneling density of states
  and the interaction correction to the conductivity, covering the
  full crossover.
\end{abstract}
\maketitle

\section{Introduction}
\label{intro}
In many regards, the theory of interacting one-dimensional (1D)
conductors seems to have reached a mature state \cite{gogolin,voit}.
In the absence of disorder, interacting 1D conductors are often
described as Luttinger liquids (LL) \cite{gogolin}.  The LL model has
been shown to be relevant for a variety of physical systems,
e.g. carbon nanotubes, semiconductor quantum wires (QWs), nanowires,
cold atoms in 1D traps, and quantum Hall edge states.  Thanks to the
work of Kane and Fisher \cite{kane}, a number of
precursors~\cite{mattis,luther,apel}, and a lot of research activity in the
past decade, the physical properties of both clean LL's and LL's
polluted by only few impurities appear to be reasonably well
understood. In particular, various expected scaling laws have been
checked numerically using the functional renormalization group
~\cite{metzner}.
 In the complementary case of significantly disordered
systems, two distinct scenarios may be realized: in quantum wires
supporting only a small number $N=\mathcal{O}(1)$ of conducting
channels, a conspiracy of interactions and impurity scattering leads
to instant localization on microscopic length scales $\sim l$, where
$l$ is the (renormalized) elastic mean free path. However, in systems
with many conducting channels --- as realized in, for instance, multi-wall
carbon nanotubes (MWNTs) --- low energy excitations may 
{\it diffuse} over distances $Nl\gg l$ much larger than the mean free
path (before Anderson localization sets in at scales $\sim Nl$.) 
The low-temperature behavior of transport
coefficients, tunneling rates, etc., is governed by strong anomalies
differing from both the characteristic power laws of the non-Fermi
liquid clean LL state, and the strongly localized disordered few-channel
 state. While for thick diffusive wires, both the
Altshuler-Aronov corrections to the conductivity \cite{aa} and the
zero-bias anomaly (ZBA) in the tunneling density of states (TDoS)
\cite{kamenev,yuli} are well
understood, much less is known about the crossover between the LL and
the diffusion-dominated regime. In one exemplary study on the 
TDoS,  Mishchenko, Andreev and Glazman
(MAG)~\cite{mishchenko} quantitatively worked out the crossover from a
LL dominated high-temperature regime to the diffusive low-energy
regime. However, the intermediate physics of several other important system
characteristics, say, the conductivity, remains unexplored.

Motivated by the fact that LL/diffusive crossover scenarios are
realized in several experimentally relevant systems, 
the purpose of this paper is the construction
of a comprehensive theory interpolating between the two regimes. We
will apply this theory to rederive earlier results on the TDoS, and
to obtain an expression of the conductivity that contains LL power laws
and Altshuler-Aronov corrections as limiting cases.
Specifically, we will formulate and analyze a model of spinless
electrons with repulsive forward-scattering Coulomb interactions and a
short-ranged Gaussian random potential.  This model will be shown to
reproduce known limiting behaviors.  Of course, in real-life systems,
additional mechanisms, such as electron-phonon scattering, magnetic
impurity scattering, or electronic (e.g.  superconducting,
charge-density wave, or Peierls) instabilities may also be important
\cite{gruner}, but we do not address those complications here.  The
theory of dirty {\it few}-channel Luttinger liquids has been pioneered
by Giamarchi and Schulz~\cite{giamarchi}, see also
Refs.~\cite{akira,ogata,sandler,nattermann,gornyi} for related
work. In contrast, we here consider the case of wide $N$-channel
wires, $N\gg 1$, termed 'quasi-1D' throughout.  We note that in the
absence of disorder, and despite the presence of many channels, LL
behavior may still be realized \cite{matv}.  This is well-known from
studies of many-chain systems \cite{gogolin,voit,gruner}, where an LL
phase is observable as long as interchain hopping can be neglected
\cite{footint}.

Besides offering various technical simplifications, the multi-channel
theory is also of considerable practical interest.  For example,
MWNTs typically carry $N\approx 5$ to $20$
open channels. Many recent MWNT experiments have shown that the
interplay of disorder and interactions plays a decisive role in
these systems~\cite{bachtold,avouris,graugnard,hakonen,kang,kanda}.
On the theoretical side, the TDoS of a
MWNT was shown to exhibit a characteristic ZBA at
low energies \cite{egger,mishchenko,kopietz}.  As far as we
know, no results for the interaction correction to the Drude
conductivity for arbitrary disorder have been published so far.
Other examples of  experimental interest include quasi-1D
organic conductors \cite{dressel}, charge-density wave
nanowires \cite{zant}, polymer nanofibers \cite{aleshin}, In$_2$O$_3$
wires \cite{liu}, or MoSe nanowires \cite{kim}.  In particular, InAs
nanowires currently receive a great deal of attention.  They have
typical mean free paths of the order 10 to 100 nm, with $N\approx 10$
to several 100 and lengths of microns up to millimetres.

Both LL power laws and the low-temperature
anomalies in diffusive systems root in the same physical mechanism:
scattering off impurities creates a $2 k_F$ oscillatory screening cloud
(Friedel oscillation) in the conduction electron density which acts as
an additional source of backscattering. 
This idea was put forward in Ref.~\cite{matveev} for a weakly
interacting 1D system with a single barrier.  For a disordered {\it
  two}-dimensional system, the ZBA~\cite{rudin} and the
conductivity~\cite{zna,gornyi1} were explored for arbitrary $T\tau_0$
within a perturbative diagrammatic framework.  However, in (quasi) 1D,
the power-law decay exponent of the Friedel oscillation itself is
affected by interactions, and the simple picture outlined above no
longer applies. Relatedly, Fermi liquid renormalization procedures
\cite{zna} must break
down in the 1D setting~\cite{gogolin}. A coherent 
description of quasi-1D systems rather 
requires the construction of a generalized approach,
and will be the main subject of this paper.
We note in passing that recent work by Golubev
 {\sl et al.}\cite{zaikin1,zaikin2}
has also explored similar crossover phenomena in 
granular models.

Building on a replicated coherent-state field integral, we will derive
an interacting nonlinear $\sigma$ model (NL$\sigma$M), similar in
spirit to that of Ref.~\cite{finkelstein}. This field theory will
allow for a nonperturbative treatment of interaction diagrams that
dominate in the ballistic limit, and thereby captures the relevant LL
physics without invoking standard bosonization
schemes~\cite{foot_boson}.  A superimposed perturbative treatment will
then allow us to compute the low-temperature diffusive corrections to
the conductivity in conceptual analogy to earlier diagrammatic
approaches.  (In many respects, the ideas entering this combination of
perturbative and nonperturbative elements are similar to those
developed in the theory of Kamenev and Andreev~\cite{kamenev}.)

The rest of the paper is organized as follows: 
In Sec.~\ref{sec2}, we describe the model and the field-theoretical
approach taken here.  
The TDoS is
discussed in Sec.~\ref{sec3}, while the linear conductivity will be
studied in Sec.~\ref{sec4}. 
Finally, in Sec.~\ref{conc}, we conclude and
offer an outlook.  Technical details have been delegated to several
appendices.  Below, we put $\hbar=k_B=1$.

\section{Model and field-theoretical approach}
\label{sec2}

\subsection{The model}

We consider a disordered and interacting quasi-1D quantum wire supporting $N$ 
conducting channels  below the Fermi energy. To describe the system we
employ the prototypical Hamiltonian
\begin{equation}
  \label{H1d}
  H=H_0 + H_{dis} + H_I,
\end{equation}
where 
$$
H_0 = -i\sum_C v_F C \int dx \Psi^\dagger \partial_x \Psi,
$$
represents the kinetic energy, $C=\pm 1$ labels right- and left-moving
states, $v_F$ is the Fermi velocity, and
$\Psi=(\Psi_{1,C},\dots,\Psi_{N,C})$ is an $N$-component vector of
field operators. Disorder scattering is described by
\begin{eqnarray}
  \label{disorder_ham}
H_{dis}&=&H_{\rm FS} + H_{\rm BS},\\ \label{hfs}
H_{\rm FS}&=&  \sum_C \int d x\, \Psi_C^\dagger(x) \hat{V}_C (x) \Psi_C
(x), \\ \label{bsd}
H_{\rm BS}  &=& \int d x\, \Psi_R^\dagger (x) \hat{W} (x) \Psi_L (x)
+ {\rm h.c.},
\end{eqnarray}
where $H_{\rm FS}$ and $H_{\rm BS}$ represent the forward (FS) and the
backward (BS) scattering, respectively. Here, $\hat{V}_C=\{V_{nn'}(x)\}$ and
$\hat{W}=\{W_{nn'}(x)\}$ are
$N\times N$ matrices in channel space which we assume to be randomly
distributed. Specifically,
\begin{equation}
  \label{dis_cor}
\langle W_{nm}(x)W^*_{m'n'}(x')\rangle_{dis}
={1\over 2\pi\nu_1\tau_0}\delta_{nn'}\delta_{mm'}\delta(x-x'),  
\end{equation}
where $\nu_1=N/\pi v_F$ is the 1D density of states, and $\tau_0$ the
(bare) elastic scattering time.  The (bare) mean free path then is
$l=v_F\tau_0$.  Finally, defining the local electron density as
$\rho=\sum_C \rho_C$, $\rho_C = \Psi_C^\dagger \Psi_C$, the
interaction Hamiltonian is defined as
\begin{equation}\label{1dints}
H_I = \frac{U}{2} \int dx  \rho^2(x),
\end{equation}
where the constant $U$ sets the interaction strength. The Hamiltonian
implies several idealizing assumptions: linearizable spectrum,
channel-independent Fermi velocity, absence of electron-electron
backscattering, and a number of others more. In Appendix \ref{1dmic} we detail
how the above Hamiltonian may be derived from a more microscopic
model, e.g. one describing a MWNT. In particular, we will argue that at 
energies $\epsilon<\epsilon_\perp$ smaller than the characteristic
energy above which structures in the direction transverse to the wire
are resolved, the above model Hamiltonian properly describes all relevant
system characteristics.

Following and generalizing Ref.~\cite{giamarchi}, a further
simplification may be effected by a gauge transformation removing the
forward scattering disorder. For $N=1$, forward scattering merely
leads to phase multiplying the electron wave functions and, therefore,
does not affect (gauge invariant) observables. A non-abelian
generalization of that argument may be used to eliminate the forward
scattering matrices $\hat{V}_c$. 
To this end, we 
perform the unitary transformation $\Psi_C (x) = \mathcal{U}_C (x)
\tilde{\Psi}_C (x),$ where
\begin{equation}\label{unitary}
\mathcal{U}_C (x) = {\cal P}
 \exp \left [ \frac{-i C}{v_F} \int^{x} d x' \hat{V}_C (x')\right ]
\end{equation}
with the path-ordering operator ${\cal P}$ defined via
\begin{eqnarray*}
&& {\cal P} \exp\left[ -\frac{iC}{v_F}
 \int^x dx' \hat V(x')\right] \equiv
\sum_{n=0}^\infty (-iC/v_F)^n \\ &&\times
 \int^x dx_n \int^{x_{n}} dx_{n-1}\cdots \int^{x_2}dx_1
\\ && \times
\hat V(x_n)\hat V(x_{n-1})  \cdots  \hat V(x_1).
\end{eqnarray*}
This unitary transformation eliminates all FS processes, $H_0+H_{\rm
  FS} \to H_0$, when $H$ is expressed in terms of the `new' fermions,
$\tilde\Psi_C$.  One can check explicitly that under this unitary
transformation, $H_{\rm BS}$ does not change its structure, and
$\hat{W}(x)$ remains a Gaussian random variable with statistical
properties \eqref{dis_cor}.  Moreover, the gauge invariant interaction
term $H_I$, and all observables studied below remain unaffected by this
transformation \cite{footgauge}. {}From now on, we switch to the
fields $\tilde\Psi$ and rename them $\Psi$.  Effectively, we can then
forget about the FS processes.

Before proceeding we note that the clean Hamiltonian $H_0+H_I$ 
describes a (multi-channel) LL
phase, described by the dimensionless Luttinger interaction parameter
$K$ and the plasmon velocity $v$,
\begin{equation}\label{kdef}
K = \frac{1}{\sqrt{1+\nu_1 U}}, \quad v= v_F/K. 
\end{equation} 
This LL phase comes along with $N-1$ (effectively noninteracting)
plasmon excitations moving at velocity $v_F$ \cite{matv}.  For $K=1$,
we retrieve a Fermi gas, while the repulsively interacting case
corresponds to $K<1$.

\subsection{Hubbard-Stratonovich transformations}
\label{seciib}

Our aim is to formulate a tractable low-energy field theory from the above
model. Restricting ourselves to equilibrium phenomena, we will employ
an imaginary time formalism and the replica trick: To compute the 
free energy $F=-T \ln Z$, we use that $\ln Z=
\lim_{r\to 0} (Z^r-1)/r$ and represent the disorder average of the
$r$th power of the partition function $\langle Z^r\rangle_{dis}$ in
terms of an $r$-fold replicated coherent-state Grassmann field
$\psi_{nC\alpha}(x,\tau)$, where $\alpha=1,\ldots,r$. (As usual, an
analytic continuation $r\to 0$ to be  performed in the end of the
calculation is implied.)  

We begin by decoupling the interaction Hamiltonian in 
a standard way via a Hubbard-Stratonovich (HS) transformation. Introducing
the (replicated) real field $\varphi_\alpha(x,\tau)$, this
generates the action for the interaction part,
\begin{equation}\label{si}
S_I = \sum_\alpha \int dx d\tau \left(\frac {\varphi_\alpha^2 }{2U} +
i \varphi_\alpha \rho_\alpha \right).
\end{equation}
Note that the fields $\varphi$ do not carry a channel index, but
couple only to the 1D density $\rho(x)$.
  (For the more general case $g_2 \neq g_4$ discussed
in Appendix \ref{1dmic} one introduces a doublet $\varphi_C$ and
proceeds in a similar way.)

One advantage of the replicated framework is that the disorder average
may be conveniently performed at an early stage of the
calculation. The action describing the disorder contribution to
$\langle Z^r\rangle_{dis}$ then assumes the form
\begin{eqnarray*}
S_{dis} &=& \frac{1}{4\pi\nu_1\tau_0}
 \sum_{C,nm,\alpha\alpha'}
 \int dx d\tau d\tau'  \\ &\times& \bar{\psi}_{nC\alpha}(x,\tau)
\psi_{n C\alpha'}(x,\tau')
\\ &\times& \bar{\psi}_{m,-C,\alpha' } (x,\tau')
\psi_{m,-C,\alpha} (x,\tau),
\end{eqnarray*}
where we used Eq.~\eqref{dis_cor}.
We decouple the time-nonlocal action $S_{dis}$
by  another Hubbard-Stratonovich transformation to obtain
\begin{eqnarray} \label{fullact}
S_{dis} & = & \frac{\pi \nu_1}{8\tau_0} \int dx d\tau d\tau' 
\sum_{\alpha\alpha'}
\tilde Q_{R,\alpha\alpha'}(x,\tau,\tau') \\ \nonumber &\times& 
\tilde Q_{L,\alpha'\alpha}  
(x,\tau',\tau) 
+\frac{i}{4\tau_0}  \int dx d\tau d\tau' \sum_{nC,\alpha\alpha'} 
\\ \nonumber
& \times & \bar{\psi}_{nC\alpha} (x,\tau) 
\tilde Q_{C,\alpha\alpha'}  (x,\tau,\tau')\psi_{nC\alpha'} (x,\tau').
\end{eqnarray}
This HS transformation introduces a functional
integral over a pair of time-bilocal but spatially local 
auxiliary matrix fields $\tilde Q_{R/L}$. To warrant convergence of 
the functional integral, the condition
\begin{equation} \label{qlqr}
\tilde Q_R^\dagger = \tilde Q_L
\end{equation}
is imposed.  Implicitly understood in the above discussion is the case
of broken time-reversal invariance, e.g., by application of a weak
magnetic field.  Otherwise, proper inclusion of the Cooperon channel
(describing, e.g., one-loop weak localization effects) would require
an additional doubling of the field degrees of
freedom~\cite{finkelstein}.  

At this stage, we may integrate out the fermion fields to arrive at the 
action 
\begin{eqnarray}\label{fullact2}
S[\tilde Q,\varphi]  &=&  \frac12 {\rm Tr} \varphi U^{-1} \varphi +
\frac{\pi \nu_1}{8\tau_0}  {\rm Tr} (\tilde Q_{R} \tilde Q_{L}) \\
\nonumber &-&  N\sum_{C=\pm}{\rm Tr} \ln \left( \partial_C
+i \varphi  + \frac{i}{4\tau_0} \tilde Q_C   \right),
\end{eqnarray}
where $\partial_C=\partial_\tau - i C v_F \partial_x$.  The trace
symbol indicates summation over replica indices and integration over
space and (imaginary) time.

In the noninteracting limit
$U=0$, a standard gradient expansion of the tracelog in Eq.\
(\ref{fullact2}) leads to a diffusive action. This expansion is
stabilized by the condition $N\gg 1$ (for $N=1$, there is of course no
diffusive phase), and starts by identifying the saddle-point solution.
It is convenient to switch from time to energy space via a double
Fourier transform. The saddle configurations are then homogeneous
($x$-independent), parity invariant ($\tilde Q_R=\tilde Q_L$), and
time-translational invariant (diagonal in energy space).  The
noninteracting saddle
$\tilde Q_{R/L} =  \Lambda$ 
can be expressed using the standard notation \cite{finkelstein}
\begin{equation} \label{lambda}
\Lambda_{\alpha\alpha^\prime}
(\epsilon,\epsilon')=  {\rm sgn} (\epsilon)
 \delta_{\epsilon\epsilon'}  \delta_{\alpha\alpha^\prime}  .
\end{equation} 
[Our Fourier convention is
$\psi(x,\tau) = T \sum_\epsilon
\int \frac{dq}{2\pi} e^{iqx + i \epsilon \tau} \psi(q,\epsilon)$,
with fermionic Matsubara frequencies $\epsilon=(2m+1)\pi T$ (integer $m$).] 
In order to study diffusion properties, we need to include fluctuation
modes around the saddle $\Lambda$.  There is one important
subtlety at that stage, which we discuss in detail in Appendix
\ref{appa}, namely there are two different types of massive
fluctuation modes.  These are (i) excitations with $\tilde Q^2 \ne 1$,
and (ii) chirally asymmetric excitations with $\tilde Q^2=1$ but
$\tilde Q_L \ne \tilde Q_R$ [involving the field $W_1$ in
Sec.~\ref{seciie}].  Excitations of type (i) are not coupled to
massless fluctuations at the Gaussian level, and hence can safely be
neglected.  However, type-(ii) excitations are linearly coupled to
massless excitations [the field $W_0$ in Sec.~\ref{seciie}], and thus
cannot simply be ignored.  In fact, only when integrating them out, we
do obtain the physically meaningful 1D result for the longitudinal
diffusion constant, $D = v_F^2 \tau_0$.  Moreover, we also recover the
transversal diffusion constant.  Including only Gaussian fluctuations
around the saddle \eqref{lambda}, we thus reproduce the full
anisotropic (cf. remark in the end of Appendix \ref{appa}) diffusive
theory \cite{egger,mishchenko}, with the correct longitudinal and
transversal diffusion constants. This provides an a posteriori
justification of the simplifying assumptions underlying the effective
Hamiltonian \eqref{H1d}.

\subsection{Chiral anomaly}

We now turn to the interacting case and look for the saddle-point
configurations corresponding to the action~\eqref{fullact2}.
Unfortunately, we are unable to determine the exact saddle, except for
the clean limit ($\tau_0\to\infty$). Hence we will adopt an {\sl
  Ansatz}, motivated by Ref.~\cite{kamenev}, which will lead to
results generally exact to lowest order in $U$, and to any order in
the ballistic limit. (In the diffusive case, the approximate saddle
point scheme is good enough to address the strongest low-temperature
singularities in transport coefficients.)  Specifically, we will, for
fixed $\varphi_\alpha$, consider field configurations of the form
\begin{equation}\label{form}
\tilde Q_C (x,\tau,\tau')= e^{i K_{C}(x,\tau)} \ \Lambda(\tau-\tau')
 \  e^{-i K_{C}(x,\tau')},
\end{equation}
i.e. configurations differing from the noninteracting saddle $\tilde
Q_C=\Lambda$ only by a chiral gauge transformation, mediated by a
space-time local field $K_C=\{K_{C\alpha}(x,\tau)\}$. Varying the
action with respect to $K_C$ determines the particular $K_C$ that
solves the saddle-point equation to lowest order in the interaction.

Before proceeding to the saddle-point solution, let us analyze 
the effect of a gauge transformation breaking chiral symmetry.
To remove the phase factors dressing $\Lambda$ in Eq.~\eqref{form},
we apply a gauge transformation to the fermionic
fields corresponding to the tracelog in Eq.~\eqref{fullact2},
\begin{equation}\label{trafo}
\psi_C (x,\tau) \mapsto e^{i K_{C} (x,\tau)} \psi_C (x,\tau).
\end{equation}
This transformation has two effects, namely (i) the term $\varphi$ in
the tracelog is replaced by $\Gamma_C$, where
\begin{equation} \label{gammac}
\Gamma_{C} =  \varphi+\partial_C K_{C},
\end{equation}
and (ii) the classical chiral $U(1)$ invariance is broken at the
quantum level, leading to a {\sl chiral anomaly} \cite{zinn,fujikawa}.
The anomaly arises from the Jacobian of the measure of the fermionic
path integral under the chiral gauge transformation \eqref{trafo}, and
is at the basis of path-integral bosonization.  The calculation can be
done by separating gauge transformations like Eq.~\eqref{trafo} into
infinitesimal steps, and then adding the corresponding Jacobian
contributions \cite{fujikawa}.  One finds that the $Q_C$ fields,
describing low-energy properties, do not contribute to the anomaly
(which is a high-energy feature).  For the sake of simplicity,
however, we here show a different derivation solely relying on known
results for the clean case.

We first recapitulate the 'standard' form of the anomaly~\cite{zinn},
\begin{equation}\label{cleananom}
- N {\rm Tr}\ln ( \partial_C +i\varphi ) = - N {\rm Tr}\ln ( \partial_C)
+ \frac{1}{2} {\rm Tr} ( \varphi \Pi_C \varphi ),
\end{equation}
where the frequency-momentum representation of the 
chiral polarization bubble is given by
\begin{equation}\label{chirbubble}
\Pi_C(q,\omega_m)= \frac{\nu_1}{2} \frac{v_F q}{v_F q+iC\omega_m}
\end{equation}
and  $\omega_m=2\pi mT$ (integer $m$) are bosonic Matsubara
frequencies. 
For later convenience, we also define
\begin{equation}\label{pol2}
\Pi (q,\omega_m) =
\sum_C \Pi_C(q,\omega_m)= \nu_1 \, \frac{(v_F q)^2}{( v_F q)^2 +  \omega_m^2}.
\end{equation}
We then use the identity
\begin{equation}
  \label{anom1}
\begin{split}
& {\rm Tr} \ln
\left( \partial_C + i \varphi + \frac{i}{4\tau_0}
\tilde{Q}_C
\right) = \\[2mm]
& {\rm Tr}\ln ( \partial_C +i\varphi ) +
 {\rm Tr} \ln
\left( 1 + \frac{i}{4\tau_0} (\partial_C + i \varphi)^{-1}  
 \tilde{Q}_C \right)=\\[2mm] 
& {\rm Tr}\ln ( \partial_C +i\varphi ) +
 {\rm Tr} \ln
\left( 1 + \frac{i}{4\tau_0} (\partial_C + i \Gamma_C)^{-1}  
 \Lambda \right),
\end{split}  
\end{equation}
where $\tilde{Q}_C$ is given by~\eqref{form},
we used
\[
 e^{-i K_{C}}  (\partial_C + i \varphi)^{-1} e^{i K_{C}} 
=  (\partial_C + i \Gamma_C)^{-1},
\]
and the cyclic invariance of the trace. (Unlike with the UV singular
logarithmic contribution to the trace, usage of 'cyclic invariance' is
a permissible operation in the expansion terms.) Similarly,
\begin{equation}
  \label{anom2}
\begin{split}
& {\rm Tr} \ln
\left( \partial_C + i \Gamma_C + \frac{i}{4\tau_0}
\Lambda
\right) = \\[2mm]
& {\rm Tr}\ln ( \partial_C +i\Gamma_C) +
 {\rm Tr} \ln
\left( 1 + \frac{i}{4\tau_0} (\partial_C + i \Gamma_C)^{-1}  
 \Lambda \right).
\end{split}  
\end{equation}
Subtracting Eqs.~\eqref{anom1} and \eqref{anom2}, and using
Eq.~\eqref{cleananom}, we obtain
\begin{equation}\label{chirano}
\begin{split}
& - N {\rm Tr} \ln
\left( \partial_C + i \varphi + \frac{i}{4\tau_0}
e^{i K_{C}} \Lambda e^{-i K_{C}}  
\right)  \\[2mm]
&=  - N {\rm Tr} \ln
\left( \partial_C + i \Gamma_C + \frac{i}{4\tau_0}
 \Lambda   
\right) \\[2mm]
& + \frac{1}{2} {\rm Tr} ( \varphi \Pi_C \varphi )
- \frac{1}{2} {\rm Tr} ( \Gamma_C \Pi_C \Gamma_C ).
\end{split}
\end{equation}
The last two terms constitute the 
chiral anomaly contribution to the action. Summing
over $C$, the anomaly reads
$S_a = \frac12 {\rm Tr}( \varphi \Pi_a \varphi)$,
with 
\begin{equation}\label{kernela}
\Pi_a  = \Pi  - \sum_C \frac{\Gamma_C }
{ \varphi} \Pi_C  \frac{\Gamma_C  }
{\varphi}.
\end{equation}
Clearly, both $\Gamma_C$ and $\Pi_a$ depend on the field $K_C$.
As a result, Eq.~\eqref{form} together with the
gauge transformation~\eqref{trafo} puts the action~\eqref{fullact2}
into the equivalent form
\begin{eqnarray}\label{startact}
 S  &=& \frac12 {\rm Tr}  \left [ \varphi \, ( U^{-1} +\Pi_a ) 
\varphi \right]  \\ \nonumber  &+&
\frac{\pi \nu_1}{8 \tau_0} {\rm Tr}\left[ e^{i K_2} \Lambda e^{-i K_2} \Lambda 
\right]  \\ \nonumber
&  - N & \sum_{C} {\rm Tr} \ln \left(\partial_C + i \Gamma_C
+ \frac{i}{4\tau_0} \Lambda \right),
\end{eqnarray}
where $K_2=K_R-K_L$.

\subsection{Saddle-point solution}\label{kam}

Our so far discussion applied to arbitrary time-dependent gauge
transformations $K_C$. We now wish to find values of $K_C$ such that the
transformed $Q$-field \eqref{form} represents an (approximate)
saddle point of the theory. To this end, we define
\begin{equation} \label{newqc}
\tilde Q_C = e^{i K_{C}} \ Q_C \  e^{-i K_{C}}
\end{equation}
and determine $K_C$ such that at $Q_C=\Lambda$ the action is
approximately stationary. Put differently, we require that at the
saddle-point values of $K_C$, first-order variations $Q_C \to \Lambda
+ \delta Q_C$ vanish. Substituting $\Lambda\to \Lambda
+ \delta Q_C$ into Eq. \eqref{startact} and expanding to linear order
in $\delta Q_C$, we thus obtain the saddle-point equations (one for
each value of $C=\pm$)
\begin{eqnarray}\label{sadl}
&&
e^{-iCK_2(x,\tau)} \Lambda(\tau-\tau') e^{iCK_2(x,\tau')}
\\ \nonumber && = \frac{2iN}{\pi\nu_1}  
\left(\partial_C+ i\Gamma_C +
\frac{i}{4\tau_0}\Lambda\right)_{x,\tau,\tau'}^{-1}.
\end{eqnarray}
These equations can generally not be solved exactly. Following
Ref.~\cite{kamenev} we will thus be content with determining a
solution that holds to leading order in the interaction strength $U$.
Expanding Eq.~(\ref{sadl}) to lowest order in $\varphi$ and hence $U$,
we find
\begin{equation}\label{linsad}
[K_2, \Lambda]_- =  \frac{2iN C}{\pi\nu_1}
\left(G_C (\varphi + \partial_C K_{C}) G_C\right)_{x,\tau,\tau'}.
\end{equation}
Here we define the noninteracting chiral disordered Green function as 
\cite{footgreen}
\begin{equation}\label{disordergr}
G_C^{-1} = \partial_C + (i/4 \tau_0) \Lambda.
\end{equation}
One may then check that 
\begin{equation}\label{saddle3}
K_{C\alpha}(q,\omega_m) =
 \frac{i \omega_m + q v_F C +  i \,{\rm sgn}
(\omega_m)/\tau_0}
{\omega_m^2+ (q v_F)^2 +  |\omega_m|/\tau_0} \, \varphi_\alpha (q,\omega_m)
\end{equation}
solves Eq.~\eqref{linsad}. In the clean (ballistic) limit, the
solution reduces to 
\[
K_C (q,\omega_m) =  -\frac{\varphi(q,\omega_m)}{i\omega_m - C v_F q}.
\] 
Indeed, these functions represent exact (to all orders in $U$)
solutions of the saddle point equation~\eqref{sadl}. In the
complementary diffusive limit $|\omega_m|\tau_0, |q| l\ll 1$,
$K_{R/L}$ asymptote to \cite{kamenev}
\[
K_{R/L}(q,\omega_m)  =\frac{i {\rm sgn}(\omega_m)}{D q^2  + |\omega_m|} 
\ \varphi (q,\omega_m),
\]
where $D=v_F^2\tau_0$ and the left-right asymmetry vanishes.  With Eq.~\eqref{saddle3}, 
$\Gamma_C$ follows from Eq.~\eqref{gammac} as
\begin{equation}\label{wxc}
\Gamma_{C}(q,\omega)  
= 2v_F q \frac{v_F q + iC(\omega+ 
{\rm sgn}(\omega)/2\tau_0)}{\omega^2+(v_F q)^2 + |\omega|/\tau_0} \
\varphi(q,\omega),
\end{equation}
and  Eq.~\eqref{kernela} yields
\begin{equation}\label{pia}
\Pi_a (q, \omega) = \nu_1 \, (v_F q)^2
 \, \, \frac{-3 (v_F q)^2 + ( |\omega| + 1/\tau_0 )^2}
{(\omega^2 + (v_F q)^2 + |\omega | /\tau_0)^2}.
\end{equation}

\subsection{Interacting non-linear sigma model}
\label{seciie}

Let us now substitute Eq.~\eqref{newqc} into the original
action~\eqref{fullact2} to obtain
\begin{eqnarray}\label{startaction}
 S  &=& \frac12 {\rm Tr}  \left [ \varphi \, ( U^{-1} +\Pi_a ) \varphi \right]  \\ \nonumber  &+&
\frac{\pi \nu_1}{8 \tau_0} {\rm Tr}\left[ e^{i K_2} Q_R e^{-i K_2} Q_L 
\right]  \\ \nonumber
&  - N & \sum_{C=\pm} {\rm Tr} \ln \left(\partial_C + i \Gamma_C
+ \frac{i}{4\tau_0} Q_C \right).
\end{eqnarray}
To make further progress, we need to include fluctuations in $Q_C$
around the saddle $\Lambda$ of the interacting system.  These
fluctuations may be conveniently parameterized as
\begin{equation}\label{param}
Q_C = T_C \Lambda T_C^{-1}, \quad T_C = \exp(-W_0/2) \exp(iC W_1/2),
\end{equation}
where $[W_{0,1},\Lambda]_+=0$ for the relevant low-energy modes.
(Fluctuations departing from the manifold \eqref{param} are massive
and decoupled from massless excitations at the Gaussian level.
Therefore they have no significant effect as long as $N \gg 1$.) The
condition $[W_{0,1},\Lambda]_+=0$ is resolved by setting
\begin{equation}\label{param2}
W_{0,1}   = \begin{pmatrix} 0 & B_{0,1} \\ -B_{0,1}^\dagger& 0 \end{pmatrix}, 
\end{equation}
where the $2\times 2$ structure refers to energy space
($\epsilon>0,\epsilon<0$).  The soft fluctuations of the theory are
thus parametrized by unconstrained complex-valued matrix elements
$(B_{0,1})_{\epsilon\epsilon^\prime,\alpha\alpha^\prime}(x)$, with
$\epsilon>0, \epsilon^\prime<0$.  Note that this parametrization gives
$Q_R^\dagger=Q_L$ as required by Eq.~\eqref{qlqr}, and
$Q_R^2=Q_L^2=1$.

At this point, 
it is convenient to integrate out high-energy fluctuations of the
field $\varphi$ within the framework of a  standard
renormalization group (RG) analysis~\cite{giamarchi}.
Starting with the bandwidth $\omega_c$ (which is of the order of the
Fermi energy) as an UV cutoff, we subsequently integrate out all
$\varphi$ fields in the strip $\omega_c'<|\omega_m|< \omega_c$. Here,
the intermediate cutoff $\omega_c'\gg (\tau_0^{-1},T)$, but small
enough to allow for a perturbative treatment of the remaining
(low-energy) interaction field.  We will show in Sec.~\ref{seciva}
that the precise value of $\omega_c'$ drops out in the calculation of
observables like the interaction correction to the conductivity.  
To implement the RG, we temporarily switch to yet another gauge
representation of the theory, viz.
\begin{eqnarray} 
\label{equivalent}
S  & = & \frac12 {\rm Tr} [ \varphi  ( U^{-1} + \Pi ) \varphi] 
+\frac{\pi \nu_1}{8\tau_0}  {\rm Tr} (e^{i K_2} Q_{R}
e^{-i K_2} Q_L )  \\ \nonumber
& -& N \sum_C  {\rm Tr} \ln \left(\partial_C + \frac{i}{4\tau_0} 
e^{i (K_{C} - K_C^0)}
Q_C  e^{-i (K_{C} - K_C^0)} \right),
\end{eqnarray}
differing from Eq.~\eqref{startaction} by a gauge transformation
mediated by $\exp(i(K_C^0-K_C))$, where $\partial_C K_C^0 = -
\varphi$. Within the high-energy strip, $B$-fluctuations,
coupling to the action through a 'low-energy' perturbation of
$\mathcal{O}(1/\tau_0)$, are expected to be of no importance, and will
be disregarded here. It then turns out that high-energy $\varphi$
fluctuations renormalize both $\tau_0$'s, the one in the second term
and that inside the tracelog of Eq.~\eqref{equivalent}, in the same
manner. Since $K_{C} - K_C^0=K_2$ for high energies, this
renormalization follows by evaluation of $\langle e^{i[K_2(x,\tau)-K_2
  (x,0)]}\rangle$ for energies within the strip, using the first term
of Eq.~\eqref{equivalent} to carry out this Gaussian average.  As net
result, we end up with an effective renormalization of the mean free
time,
\begin{equation}\label{tau0} 
\tau_0 \to \tau_0' = \tau_0 (\omega'_c/\omega_c)^{\gamma_i},
\end{equation}
with the Luttinger liquid power-law exponent for weak backscattering
by a single impurity \cite{kane}, 
\begin{equation}\label{gi}
\gamma_i = 2 (1-K)/N,
\end{equation}
where $K$ is the Luttinger parameter \eqref{kdef}.
In what follows, we imagine that the RG has been carried out,
and only discuss its effects explicitly where required. 
(In Sec.~\ref{sec3}, it will in fact be more convenient not to perform
the RG transformation at all.)

Its logarithmic nonlinearity makes the general analysis
of the action~\eqref{startaction} difficult. For momenta $q\lesssim
1/(v_F \tau_0')$, 
and frequencies $\omega \lesssim \tau_0^{\prime -1}$, 
probing the multiple scattering
regime, a gradient expansion in small momenta can be
performed to obtain 
\begin{eqnarray}\label{nlsm}
S &=&\frac12 {\rm Tr}  \left [ \varphi \, ( U^{-1} +\Pi_a ) 
\varphi \right]  \\ \nonumber  &+&
\frac{\pi \nu_1}{8 \tau_0} {\rm Tr}\left[ T_R \Lambda T_R^{-1} e^{-i K_2}
T_L \Lambda T_L^{-1} e^{i K_2}
\right]  \\ \nonumber
&+& \frac{i \pi \nu_1}{2}   \sum_{C} {\rm Tr} \left( \Lambda ( T_C^{-1} 
\partial_C T_C + i \Gamma_C) \right) \\ \nonumber
&+& \frac{\pi \nu_1 v_F}{2} \sum_C {\rm Tr} \left( G_C (T_C^{-1} \partial_C T_C 
+i \Gamma_C ) \right)^2+\dots
\end{eqnarray}
where the Green functions are given by \eqref{disordergr}. In Appendix~\ref{appa}
we show how the standard form of the diffusive
 $\sigma$-model may be obtained from this prototype action.
However, for arbitrary momenta and frequencies no such
reduction is possible. Nonetheless progress can be made,
viz.~by noting that an expansion of the action \eqref{startaction} to second
order in the fluctuation generators $B$ is sufficient to compute
our observables of prime interest. In essence, this expansion is
tantamount to an RPA approximation to the model where the 'RPA-bubble'
accounts for impurity scattering. 
Our neglect of higher-order fluctuations 
essentially implies that the renormalization
of diffusion and interaction by quantum
interference processes is not taken into account.

\section{Zero-bias anomaly}
\label{sec3}

To warm up, before addressing the more involved interaction
corrections to the conductivity in Sec.~\ref{sec4}, let us briefly
discuss the energy-dependent bulk tunneling density of states (TDoS),
see also Refs.~\cite{egger,mishchenko,kopietz}.  The TDoS governs
temperature- or voltage-dependent transport quantities when contacts
or other conductors are weakly coupled to the QW.  It is well known
that in both limits, Luttinger and diffusive, the DoS exhibits
pronounced structure as a function of energy. The crossover between
the two limiting scenarios was addressed  by
MAG \cite{mishchenko} by a direct analysis of 
the screened interaction in multi-channel
quantum wires. We here rederive and extend their results, 
essentially to illustrate the application of the present formalism. 

For a particle-hole symmetric system, the TDoS is \cite{gogolin}
\begin{equation}
\rho (\omega,T) = -(2/\pi) \coth (\omega/2T) 
\sum_C \int_0^{\infty} dt \sin \omega t \, \, {\rm Im}
{\cal G}_C (t),
\end{equation}
where ${\cal G}_C(t)$ is the real time analytic continuation, $\tau\to
i t+0^+$,   of 
\begin{eqnarray}\label{ac}
&& \sum_n
\langle T_\tau \psi_{nC} (x,\tau)  \bar{\psi}_{nC} (x,0) \rangle
= \\ \nonumber && = -i\pi \nu_1 \Lambda(\tau) 
\langle e^{i[K_C(x,\tau)-K_C
(x,0)]}\rangle_{S_\varphi},
\end{eqnarray}
$T_\tau$ the time-ordering operator, and $x$ an (arbitrary) space coordinate.
Since ${\cal G}_R = {\cal G}_L$, we are free to focus on ${\cal G}_R$.  In Eq.\
(\ref{ac}), fluctuations $T_C\ne 1$ do not contribute, and the average
in Eq.~\eqref{ac} is taken using the action $S_\varphi$ obtained from
Eq.~\eqref{startaction} with $T_C=1$.  (The replica limit is then
trivial to take.)  For noninteracting fermions, $K_{C} = 0$, and hence
\begin{eqnarray*}
{\cal G}_R(t) & =& -i \pi\nu_1 \Lambda(t) = -i \pi \nu_1 T/\sinh (\pi T t),
\\ \rho (\omega,T) &=& 2\nu_1 T \coth(\omega/2T) 
\int_0^{\infty} dt \, \frac{\sin (\omega t)}{\sinh(\pi T t)}
= \nu_1.
\end{eqnarray*}
In order to analytically evaluate 
$\langle e^{i[K_R(x,\tau)-K_R
(x,0)]}\rangle_{S_\varphi}$, 
we expand $S_\varphi$ to quadratic order in $\varphi$, leading to
\begin{equation} \label{actphi2}
S_\varphi = \frac12 {\rm Tr} \varphi \Pi_0 \varphi.
\end{equation}
This truncation does not introduce any approximation 
in the ballistic limit, and is consistent with the weak-interaction
condition inherent in the construction of the saddle \eqref{form}. 
The kernel $\Pi_0$ includes (i) the term $(U^{-1}+\Pi_a)$, (ii) a 
contribution from the second term in Eq.~\eqref{startaction}, 
and (iii) from the expansion of the tracelog, a term $\sim
{\rm Tr}(G_C \Gamma_C)^2$.  For instance, (ii) gives
\begin{eqnarray*}
&&\frac{\pi\nu_1}{8\tau_0} {\rm Tr}(\Lambda K_2\Lambda K_2-K_2^2)\\
&&= {\rm Tr} \left(\varphi(-q,-\omega)
\frac{\nu_1(v_F q)^2 |\omega|/2\tau_0}{(\omega^2+(v_F q)^2+|\omega|/\tau_0)^2}
\varphi(q,\omega)\right).
\end{eqnarray*}
Combining all three terms, the kernel in Eq.~\eqref{actphi2} reads
\begin{equation}\label{crossover}
\Pi_0 (q,\omega_m) =  \frac{1}{U} + \nu_1 \, \frac{ ( v_F q)^2}
{\omega_m^2 + (v_F q)^2 + |\omega_m|/\tau_0}.
\end{equation}
This expression bridges the ballistic and diffusive
regimes. It is exact in the LL limit, while 
it describes diffusively screened interactions in the opposite
limit, assuming weak interactions.  

It is then straightforward to arrive at the
TDoS.  With the definition (\ref{kdef}) of the LL parameter $K$,
and the bulk ($\gamma$) and boundary 
($\gamma_b$) Luttinger tunneling exponents  \cite{matv},
\begin{equation}\label{exponents}
\gamma = (K+1/K-2)/2N, \quad \gamma_b = (1/K-1)/N ,
\end{equation}
some algebra leads to \cite{foot1},
\begin{eqnarray}\label{crossdos}
\frac{\rho (\omega,T) } {\nu_1} &  = &
\frac{2 }{\pi} \coth(\omega/2 T)
\, {\rm Re} \int_0^{\infty} \frac{dt}{t}
 \sin (\omega t)  \\
\nonumber &\times& \left( \frac{\pi T t}{\sinh(\pi T t)}
\right)^{\gamma+1}   (1+i \omega_c t)^{-\gamma} \\
\nonumber &\times& e^{ \gamma  F_1 (t/\tau_0,T)
 +\gamma_b H_1(t/\tau_0,T) + i \gamma F_2(t/\tau_0)+ 
i \gamma_b H_2 (t/\tau_0)},
\end{eqnarray}
where we have defined 
\begin{eqnarray*}
F_1 (y,T)  &=& {\rm Re} \int_0^{\infty} du  \left(\frac{1}{u} -
\frac{1}{\sqrt{u^2+i u}} \right) \frac{ 1- \cos (2 yu) }
{\tanh(u/ (2T \tau_0))} \\
F_2 (y) & = & \frac{\pi}{2} \left( 1- I_0(y) e^{-y} \right ) \\
H_1 (y,T)  &=& \frac{{\rm Im}}{2} \int_0^{\infty} du  
\frac{ u+i }{ (u^2+i u)^{3/2} } \frac{ 1- \cos (2 yu) }
{\tanh(u/(2T \tau_0))} \\
H_2 (y) & = & -\frac{\pi}{2} y e^{-y} \left( I_1(y) +I_0(y) 
\right) ,
\end{eqnarray*}
and $I_{0,1}$ are Bessel functions of imaginary argument \cite{gradsteyn}.
For $T=0$, the integrals for  
$F_1$ and $H_1$ can be evaluated in closed form,
\begin{eqnarray*}
F_1 (y,0) & =& - \ln \left( \frac{2 e^{-{\rm C}}}{y} \right) + K_0 (y)
\cosh(y), \\
H_1 (y,0) & = & 1-  y \left[ \cosh(y) K_1(y) +
\sinh(y) K_0 (y) \right],
\end{eqnarray*}
where ${\rm C}=0.577\ldots$ is the Euler constant, and $K_{0,1}$ are 
related Bessel functions of imaginary argument \cite{gradsteyn}.
Note again that Eq.~(\ref{crossdos}) has been derived under
the assumption of large channel number $N$.

Before proceeding, let us pause to briefly comment on the relation of 
Eq.~\eqref{crossdos} to the results of MAG.
The far reaching equivalence between Eq. (\ref{crossdos}) and MAG's
Eq.~(14) is best exposed by shifting their integration variable $t$
according to
 $t\to t-i/2T$. The results then turn out to be  identical,
 provided one identifies the integral kernel $\mathcal{V}(\omega)$ of
 MAG with
\[
{\cal V}(\omega)= -\frac{1/K-1}{2N} {\rm Re}
\frac{\omega(1-K)+i/\tau_0}{\omega^{3/2} \sqrt{\omega+i/\tau_0}}.
\]
Effectively, however, in MAG the limiting case of 
strong interactions and/or large
channel number, $K\to 0$, is considered. In the ballistic regime, and
for moderate interactions and channel numbers, their results
do, therefore, not scale according to the Luttinger exponent
Eq.~(\ref{exponents}). Within the framework of MAG's analysis, this
deviation is however straightforward to avoid by not taking the limit of
strong interactions at an early stage.

\begin{figure}[t!]
\vspace{0.2cm}
\scalebox{0.3}{\includegraphics{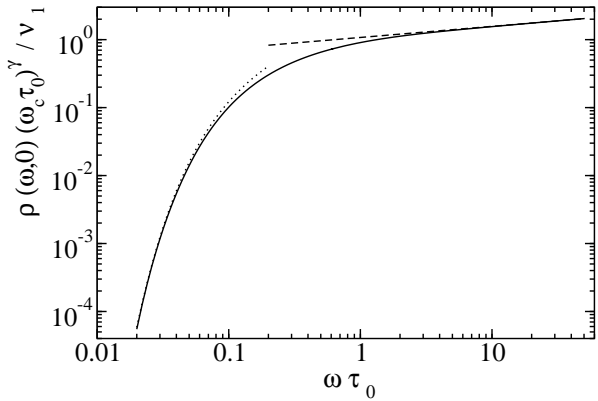}}
\caption{ \label{fig1} 
Zero-temperature crossover solution (\ref{crossdos}) for the TDoS (full curve)
at $K=0.3$ and $N=5$.  Here $\omega_c\tau_0=5\times 10^4$ was chosen, which is 
large enough to render results insensitive to its value.
Dotted and dashed curves give the limiting
expressions (\ref{diffdos}) and (\ref{tdosll}), respectively.
Note the double-logarithmic scales. }
\end{figure}

We next discuss Eq.~\eqref{crossdos} in various limits of
interest. Beginning with the zero temperature case, $T=0$, for
$\omega\tau_0\gg 1$, only small $t$ contribute to the integral
(\ref{crossdos}). The functions $F_{1,2}$ and $H_{1,2}$ are then
basically negligible, the $t$-integral can be done, and we obtain the
well-known LL power-law TDoS \cite{gogolin},
\begin{equation} \label{tdosll}
\frac{\rho(\omega)}{\nu_1} = \frac{1}{\Gamma(1+\gamma)} 
(\omega/\omega_c)^\gamma
\end{equation}
with the Gamma function $\Gamma(x)$.  In the diffusive regime,
$\omega\tau_0\ll 1$, 
large $t$ dominate in
Eq.~(\ref{crossdos}), and one can use asymptotic expansions for
$F_{1,2}$ and $H_{1,2}$.  
Actually, the precise crossover scale
separating the diffusive from the ballistic regime in the $T=0$
TDoS is set by $\omega\tau_0\sim \gamma$, see Ref.~\cite{mishchenko}. 
With the error function $\Phi$, we find
\begin{eqnarray} \nonumber 
\frac{\rho(\omega)}{\nu_1} & = &
\frac{ e^{ {\rm C}\gamma+\gamma_b} }{(2\omega_c\tau_0)^\gamma} 
\left[ 1- \Phi \left( \sqrt{\frac{2\pi}{\omega\tau_0}} \ \gamma_b\right) \right]
 \\ &\simeq& \label{diffdos} 
\frac{ e^{{\rm C}\gamma+\gamma_b} }{(2\omega_c\tau_0)^\gamma} 
\frac{\sqrt{\omega\tau_0/2}}{\pi \gamma_b}
 e^{-2\pi\gamma_b^2/(\omega\tau_0)},
\end{eqnarray}
corresponding to an exponentially vanishing TDoS at low
energies. This pseudogap behavior was first noted  by Nazarov \cite{yuli},
 and subsequently rederived in Refs.~\cite{mishchenko,kopietz}.
While a direct expansion of the $\sin(\omega t)$ in Eq.~(\ref{crossdos})
would suggest a linearly vanishing TDoS \cite{kopietz}, the correct result
is the nonperturbative exponential law (\ref{diffdos}).  In fact,
one can show by complex integration methods that in the 1D case
the prefactor of any polynomial-in-$\omega$ term must vanish.
Similarly, one can show that the $T$-dependence of 
$\rho(\omega=0)$ also has to be exponential as $T\to 0$.
Figure \ref{fig1} shows the full $T=0$ crossover solution together
with the asymptotic results (\ref{tdosll}) and (\ref{diffdos}).

Next we turn to the regime $T\tau_0 \ll 1$ and $\omega\tau_0 \ll 1$.
The crossover function describing the transition from $\omega < T$ to
$\omega > T$, reported in Ref.~\cite{mishchenko}, can be reproduced
from Eq.~(\ref{crossdos}) using asympotic expansions of $F_{1,2}$ and
$H_{1,2}$.  In particular, for $\omega\ll T\ll 1/\tau_0$,
\begin{equation}
\frac{\rho(T)}{\nu_1}  = 
\frac{ e^{{\rm C}\gamma+\gamma_b} }{(2\omega_c\tau_0)^\gamma} 
\sqrt{\frac{\pi}{2.389}} 
\left(\frac{ T \tau_0}{2\pi \gamma_b^2} \right)^{1/4}  \, 
e^{-1.075 \gamma_b \sqrt{2\pi/ T \tau_0}}.
\end{equation}

Finally, once either $T\tau_0\gg 1$ or $\omega\tau_0 \gg 1$,
typical values for $t$ in Eq.~(\ref{crossdos}) are 
$t\sim {\rm min}(1/T, 1/\omega)$, such 
that effectively the standard LL power laws are recovered.
For $T\gg (1/\tau_0, \omega)$, 
\begin{equation}\label{asympt4}
\frac{\rho(T)} {\nu_1}  
\simeq \left( \frac{\pi}{2} \right)^\gamma \frac{\Gamma(1+\gamma)}
{\Gamma^2(1+\gamma/2)} (T/\omega_c)^\gamma.
\end{equation}

\section{Conductivity}
\label{sec4}

\subsection{Interaction correction}
\label{seciva}

In this section, we discuss the interaction correction to the Drude
conductivity for arbitrary disorder, connecting the diffusive
Altshuler-Aronov result \cite{aa} with the ballistic limit, where
Luttinger liquid power laws emerge. It is worth reiterating that weak
localization corrections are not included in our theory, see
Sec.~\ref{seciib}.

The Kubo formula determines the linear dc 
conductivity as the $q\to 0, i\omega_m \to \omega+i0^+$ limit of 
\begin{equation}
\sigma (q , \omega_m) =  -\frac{e^2}{\omega_m} \langle j (q,\omega_m)
j (-q,-\omega_m)\rangle +\frac{ e^2 \rho_0}{m \omega_m},
\end{equation}
where $\rho_0$ is the Drude resistivity, $m$ the
effective mass, and $j$ the particle current.
The conductivity can alternatively be expressed as
\begin{equation} \label{cond1}
\sigma (q , \omega_m) = - (e/q)^2 \, \omega_m \, 
P(q,\omega_m), 
\end{equation}
where the density-density correlator can be written as
\begin{equation}\label{densityfluc}
P (q,\omega) = \frac{\langle \varphi
 (q,\omega) \varphi(-q,-\omega) \rangle}{U^2} - 
\frac{1}{U}.
\end{equation}
Starting from the full action in the original formulation,
see Eq.~\eqref{fullact2}, the formula
\eqref{densityfluc} may be derived as follows:
Introduce a source field $\eta(q,\omega_m)$ that is
added to $\varphi$ in the tracelog only. The density correlator
then follows from the resulting generating functional $Z[\eta]$ by 
means of 
\begin{equation}\label{genfunc}
P(q,\omega_m)= \left(\frac{\partial^2 \ln Z}{\partial \eta(q,\omega_m)\partial
\eta(-q,-\omega_m)}\right)_{\eta=0}.
\end{equation}
Under the {\sl local} gauge transformation \eqref{trafo}, 
$\eta$ remains invariant, and one can safely perform the shift
$\varphi\to \varphi-\eta$ in the action. After that shift, 
it is a simple matter
to carry out the derivatives in Eq.~\eqref{genfunc} 
and to arrive at Eq.~\eqref{densityfluc}.
In what follows, we use the intermediate cutoff $\omega_c'$  
and the mean free time $\tau'_0=\tau_0 (\omega_c'/\omega_c)^{\gamma_i}$
with the exponent \eqref{gi}, see Sec.~\ref{seciie}.
After the corresponding RG procedure,
the remaining interactions are treated by
a first-order perturbation theory approach.

The {\sl classical  Drude conductivity} can now be recovered by
our formalism as follows.  Dropping all $B$ fluctuations
in Eq.~\eqref{nlsm}, the action up to quadratic order in $\varphi$ is
$S_\varphi$, given by Eqs.~\eqref{actphi2} and \eqref{crossover}.
Keeping only $S_\varphi$, we may then apply 
Eq.~(\ref{cond1}), continue to real frequencies and arrive at the  Drude conductivity
\begin{equation}\label{drude}
\sigma'_D = e^2 \nu_1 v^2_F \tau'_0 = \frac{Ne^2 v_F}{\pi \hbar} \tau'_0.
\end{equation}
This result includes an interaction correction encoded in the
high-frequency RG scheme above. When substituting $\tau'_0\to \tau_0$,
one recovers the standard (noninteracting) Drude conductivity,
\[
\sigma_D = e^2 \nu_1 v_F^2 \tau_0.
\] 
We mention in passing that the above approximation also obtains the
correct ac Drude form of the optical conductivity
\begin{equation}
\sigma_D(\omega)= \sigma_D/(1-i\omega \tau_0).
\end{equation}

{\sl Interaction corrections} to the dc Drude conductivity
\eqref{drude}, defined via $\sigma=\sigma'_D+\delta \sigma'$, are
obtained by inclusion of (i) $B$ fluctuations, and (ii) higher-order
contributions in $\varphi$ to the action. To lowest order in the
interactions, these additional ingredients to the theory conspire to
obtain the result
\begin{eqnarray}\nonumber
\frac{\sigma(T)}{\sigma_D}  & = & (\omega_c\tau_0)^{-\gamma_i/(1+\gamma_i)}
\Biggl[ 1 + \gamma_i \ln( 2\pi e^{-1-{\rm C}} T\tau_0'  ) \\ \nonumber
&-&\gamma_i {\rm Re}   
\int_0^{\infty} d\Omega \frac{F(\Omega)}{\Omega} \Biggl (  
\sqrt{ \frac{\Omega}{\Omega+i/\tau'_0} } -1 \\   &+&
   \frac{i}{2 \tau'_0 (1+K) \sqrt{\Omega^2 +i \Omega/\tau'_0}} \Biggr)
\Biggr],
\label{intcorr2} 
\end{eqnarray}
with the function 
\begin{equation}\label{fo}
F(\Omega) = \frac{\partial \left(\Omega {\rm coth}
(\Omega/2 T)\right)}{\partial\Omega}.
\end{equation}
For weak interactions, it is justified to use the
renormalized mean free time in the form
\begin{equation}\label{taup}
\tau_0'=\tau_0 (\omega_c\tau_0)^{-\gamma_i/(\gamma_i+1)}.
\end{equation}
Before turning to the derivation of this result, let us discuss its
physical meaning. We first note that, up to the prefactor
$(\omega_c\tau_0)^{-\gamma_i/(\gamma_i+1)}$, the conductivity is a
universal function of the interaction parameter $K$, the channel
number $N$, and the parameter $T\tau_0'$. Importantly, the crossover
temperature separating the LL and the Altshuler-Aronov regions,
respectively, is determined by the renormalized scale $\tau_0^\prime$.
For $T\tau_0'> 1$ and $\gamma\ll 1$
Eq.~\eqref{intcorr2} simplifies to
\begin{equation}
  \label{eq:1}
 { \sigma(T)\over \sigma_D}\simeq (\omega_c \tau_0')^{-\gamma_i}
\left[1+\gamma_i\ln (2\pi e^{-(1+C)} T\tau_0')\right],\;\; T\gtrsim
\tau^{\prime -1}. 
\end{equation}
Exponentiation of the logarithm obtains  the familiar Luttinger
liquid high-temperature power law \cite{mattis,giamarchi,gornyi},
\begin{equation}\label{finalhigh}
\frac{\sigma(T)}{\sigma_D} \simeq 
e^{-\gamma_i(1+{\rm C})} (2 \pi T /\omega_c )^{\gamma_i},\;\; T\gg
\tau^{\prime -1},
\end{equation}
otherwise obtained by stopping the RG procedure of Sec.~\ref{seciie}
at $T\gg \tau^{\prime -1}$. Governed by the single-impurity
backscattering exponent $\gamma_i$ (cf. Eq.~\eqref{gi} and
Ref.~\cite{kane}), the exponentiated form is  valid only at
asymptotically large temperatures where multiple interference is
negligible. 
In
the complementary regime 
$T\tau_0' \ll 1$ the interaction correction becomes
\begin{equation}\label{finallow}
\frac{\sigma-\sigma_D}{\sigma_D}  \simeq - \gamma_i
 \frac{3  \sqrt{\pi/2} \ \zeta(-1/2)}{1+K}  (T \tau^\prime_0)^{-1/2}, 
\end{equation}
where $\zeta$ is the Riemann zeta function \cite{gradsteyn}.
This expression recovers the well-known $T^{-1/2}$
Altshuler-Aronov correction in 1D \cite{aa}, leading to a pronounced
suppression of the conductivity at low temperatures.
At very low temperatures, this correction becomes sizeable and our
first-order perturbative approach ceases to be  valid.
In that case, also quantum interference corrections have to be included.

\begin{figure}[t!]
\begin{center}
\includegraphics[scale=0.3]{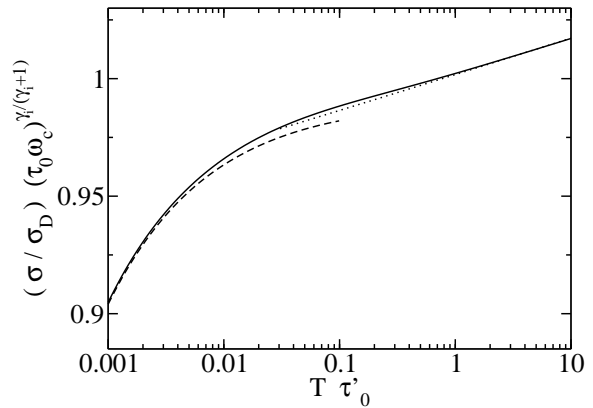}
\caption[]{\label{f2} 
Temperature dependence of the conductivity \eqref{intcorr2}
for $K=0.9$ and $N=30$. The limiting behaviors \eqref{finalhigh} and 
\eqref{finallow} 
are shown as dotted and dashed curves, respectively.}
\end{center}
\end{figure}

The crossover solution \eqref{intcorr2} smoothly interpolates between
the limits (\ref{finalhigh}) and (\ref{finallow}), respectively, as is shown 
for $K=0.9$ and $N=30$ in Figure \ref{f2}.
Note that Eqs.~\eqref{finalhigh} and \eqref{finallow} 
match each other at $T\tau_0^\prime$ of order unity.
Related crossover scenarios from diffusive  dynamics at 
low temperatures to non-diffusive dynamics at high 
temperature have been discussed in earlier work. 
Specifically, in  Refs.~\cite{zaikin1,zaikin2},
the conductance of a quantum dot array was considered. 
In  that system the high-temperature regime is governed by 
the chaotic dynamics of individual quantum dots. 
The latter differs strikingly  from the ballistic 
dynamics of clean quantum wires which is why the 
results obtained for the high-temperature conductivity
corrections  are {\it quantitatively} different from ours. 
(Nonetheless, the  qualitative high-$T$ profile looks similar, 
cf.~Fig.~2 of Ref.~\cite{zaikin1}.)

We finally note that our calculation above ignores 
finite-size effects, which are of minor importance in
sufficiently long wires with good contact to external leads,
where the resistance is dominated by the intrinsic part.
Note also that four-terminal resistance measurements allow to circumvent 
boundary effects as long as the electrodes act as non-invasive probes.
In both cases, the conductivity is the relevant transport
coefficient.  This regime has been experimentally studied in 
carbon nanotubes and other quantum wires for many years by now. 
In particular, linear voltage drops along a MWNT can be measured
and have been reported in Ref.~\cite{afm}.

\subsection{Derivation}\label{secivb}

Let us now sketch the derivation of Eq.~\eqref{intcorr2}. 
For notational simplicity,  the primes in $\tau'_0$ and $\omega_c'$
will be omitted for the moment.
Our strategy is as follows: 
We expand Eq.~(\ref{startaction}) to second order in $B$-fluctuations,
and integrate them out. This will generate an effective action for
$\varphi$, which we expand up to fourth order.  In addition, for 
internal consistency,
the $B$-independent terms in Eq.~\eqref{startaction} must also be 
kept up to fourth order.  Schematically, our starting action then reads
\begin{equation}\label{actionexpansion}
S = S_\varphi  + S_B + S_4^{(1)} + S_4^{(2)}.
\end{equation}
The interaction correction results from the last three terms.  The
contribution $S_B$ combines (i) terms of second order in $B_{0,1}$ but
zeroth order in $\varphi$, and (ii) of first order in $B$ but
second order in $\varphi$.  Note that the first order in both
$B$ and $\varphi$ is strictly zero, due to our choice of the
saddle point.  Moreover, contributions to second order in $B$
and first or second order in $\varphi$ identically vanish in the
replica limit.  Finally, 
\begin{equation}\label{act41}
S_4^{(1)} = \frac{\pi \nu_1 v_F}{4} \sum_C {\rm Tr}  ( G_C \Gamma_C )^4
\end{equation}
comes from $B$-independent terms in the expansion of the tracelog
in Eq.~\eqref{startaction}, while
\begin{equation}\label{act42}
S_4^{(2)} = \frac{\pi \nu_1}{8 \tau_0} \left( \frac{1}{12} {\rm Tr} K_2^4
- \frac{1}{3}  {\rm Tr} \Lambda K_2 \Lambda K_2^3 +   \frac{1}{4} 
{\rm Tr} \Lambda K_2^2 \Lambda K_2^2 \right)
\end{equation}
similarly follows from the second term in Eq.~\eqref{startaction}.

Since Eq.~\eqref{densityfluc} involves only $\varphi$, the $B$-fluctuations
can be integrated out at the level of the action,
replacing $S_B$ effectively by a new fourth-order term
$S_4^{(3)}[\varphi]$.  The lengthy but straightforward derivation of
this term is detailed in Appendix \ref{appb}.  Defining $S_4 \equiv S_4^{(1)}+
S_4^{(2)} + S_4^{(3)}$, the lowest-order correction to the
conductivity, Eq.~\eqref{densityfluc}, is obtained by first-order
expansion in $S_4$,
\begin{equation}\label{corr1}
\delta P(q,\omega)=-U^{-2} \left\langle \varphi(q,\omega)
\varphi(-q,-\omega) [S_4 - \langle S_4 \rangle_{S_\varphi}]
\right\rangle_{S_\varphi}.
\end{equation}
Loosely speaking, fluctuations of the $\varphi$ field ($
S^{(1)}_4,S^{(2)}_4$) describe the RPA screening of the Coulomb
interaction by impurity-dressed Green functions ('empty bubble'), while
joint fluctuations of the $B$ and the $\varphi$ field ($
S^{(3)}_4$) account for the vertex renormalization of the RPA bubbles
by impurity ladders. Accordingly, we will find that $S^{(3)}_4$ holds
responsible for the corrections to the conductivity showing the strongest
low-temperature singularities. Conversely, the terms
$S^{(1)}_4,S^{(2)}_4$ dominantly contribute to the conductivity in the
Luttinger limit $T\tau_0^{\prime}>1$.

\begin{figure}
\begin{center}
\includegraphics[scale=0.6]{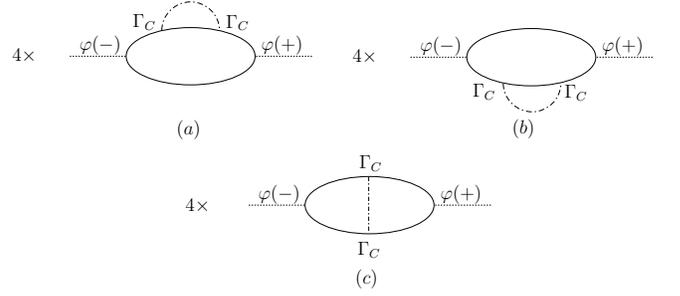}
\caption{\label{diag1} 
Diagrams corresponding to $S_4^{(1)}$ and $S_4^{(2)}$ in the conductivity
calculation. The contraction
of  $S_4^{(1)}$ or $S_4^{(2)}$ with  $\varphi(q,\omega)\varphi(-q,-\omega)$
(denoted by $\varphi(\pm)$, respectively) leads to the
pairing of $\varphi(\pm)$ with two different $\Gamma_C$ (or $K_2$ 
in the case of $S_4^{(2)}$),  implying $12$ different 
contractions. Due to the cyclic invariance of the trace, 
only three different contributions emerge, depending on the 
$\Gamma_C$ ordering.
Solid curves represent schematically the trace, external dotted lines 
the two $\Gamma_C$ connected to $\varphi(\pm)$, and dashed-dotted lines
the pairing between the remaining $\Gamma_C$. }
\end{center}
\end{figure}

Doing the $\varphi$-contractions, we obtain for the conductivity
\begin{eqnarray} \label{conduct_total}
  \delta \sigma & = &- \sigma_D \, {\rm Im} \int \frac{ d \Omega}{\pi} 
F(\Omega) \int \frac{ d q}{2\pi}
\Pi_{0,R}^{-1}(q,\Omega)  \\ \nonumber &\times& (v_F q)^2
{(K_1+K_2+K_3) (q,\Omega) \over(v_F^2 q^2 -\Omega^2 -i\Omega/\tau_0)^2},
\end{eqnarray}
with the auxiliary quantities
\begin{eqnarray}
K_1(q,\Omega)&=&4\Bigl( 1 + \frac{1-2i\Omega\tau_0}{(2v_F q\tau_0)^2 + (1-2i\Omega_0\tau_0)^2)} 
\Bigr),\nonumber\\
K_2(q,\Omega)&=&-2,\nonumber\\
K_3(q,\Omega)&=&\frac{1}{\tau_0^2 (v_F^2 q^2 -\Omega^2 -i\Omega/\tau_0)}\times \\
\nonumber &\times& \frac{2-6i\Omega\tau_0-(2\Omega\tau_0)^2 + 
(2v_F q\tau_0)^2} 
{(1-2i\Omega\tau_0)^2+(2v_F q\tau_0)^2},
\end{eqnarray}
where the function $F(\Omega)$ has been defined in (\ref{fo}). 
The contributions $K_{1,2,3}$ stem from $S_4^{(1,2,3)}$,
respectively, and $\Pi_{0,R}^{-1}$ denotes the retarded function
corresponding to $\Pi_0^{-1}$, see
Eq.~\eqref{crossover}. Specifically, the 'ballistic contributions'
$K_{1,2}$ result from summing 12 terms corresponding to the diagrams
in Fig.~\ref{diag1}, followed by analytic continuation to real
frequencies. (Although standard in principle, the actual calculation
obtaining these terms, in particular the analytic
continuation to real frequencies for the product of four Green
functions, is rather involved. We refer to Appendix A of
Ref.~\cite{zna} for related technical details.) The calculation of the
third term $K_3$ is detailed in Appendix \ref{appb}. Dominating in the
diffusive limit, it obtains the standard 1D Altshuler-Aronov
correction \cite{aa}, see Eq.~\eqref{finallow}.

Equation \eqref{conduct_total} implies  the total correction  to
the Drude conductivity to lowest order in the interaction,
\begin{eqnarray}\label{cross-conduc}
\delta \sigma &=&  - 2 \sigma_D U \, \, {\rm Im} 
\int \frac{ d \Omega}{\pi} F(\Omega)
\int \frac{ d q}{2\pi} \\ \nonumber &\times&
\frac{v_F^2 q^2 \left[v_F^2 q^2 + (1/\tau_0-i\Omega)^2\right]}{(v_F^2 q^2-\Omega^2-i\Omega/\tau_0)^2
(v^2 q^2-\Omega^2-i\Omega/\tau_0)},
\end{eqnarray} 
where $v=v_F/K$, see Eq.~\eqref{kdef}.
After carrying out the final momentum integration
and restoring primed quantities, we arrive at
 the preliminary result
\begin{eqnarray}\label{intcorr} 
\delta \sigma'  & = & - \gamma_i \sigma'_D {\rm Re}   
\int_0^{\omega_c'} d \Omega \frac{F(\Omega)}{\Omega} \Biggl[  
\sqrt{ \frac{\Omega}{\Omega+i/\tau'_0} }  \\  \nonumber &+&
   \frac{i}{2 \tau'_0 (1+K) \sqrt{\Omega^2 +i \Omega/\tau'_0}} \Biggr],
\end{eqnarray}
which may give the misleading impression that 
the conductivity depends on the somewhat arbitrary cutoff $\omega_c'$.
Noting that $\omega_c'\tau_0'\agt 1$,
 to first order in the interaction, we have 
\[
\tau'_0 = \frac{\tau_0}{(\omega_c \tau_0)^{\gamma_i/(\gamma_i+1)}} 
\left( 1+ \gamma_i \ln (\tau'_0 \omega_c') \right),
\]
leading to the estimate (\ref{taup}).
Splitting off the UV divergent part in Eq.~\eqref{intcorr}
leads to the logarithmic term in Eq.~\eqref{intcorr2}, 
while the upper limit  can 
be sent to infinity for the remaining  integral in
Eq.~\eqref{intcorr}. We finally arrive at the
$\omega_c'$-independent form (\ref{intcorr2}).

\section{Conclusion}
\label{conc}

In this paper, we have developed a low-energy field theory 
of weakly interacting disordered multi-channel conductors. 
The theory is formulated in terms of  two auxiliary fields, 
a scalar field $\varphi(x,\tau)$ decoupling the electron-electron interactions,
and  a matrix field $Q_C(x,\tau,\tau')$ decoupling the 
effective interactions arising from the disorder ensemble average. The
two fields $\varphi$ and $Q_C$ are coupled by a gauge mechanism. 
For general values of interaction/disorder/channel number, the theory
is governed by a nonlinear action --- the notorious 'tracelog' --- and
remains difficult to evaluate. However, in a number of important
cases analytical progress is possible. Specifically, at energies
$\epsilon<\tau_0^{-1}$ smaller than the elastic scattering rate, and for
large channel numbers, we recover the diffusive interacting
$\sigma$ model\cite{finkelstein} with all the known consequences. In
the opposite limit $\epsilon \gg \tau_0^{-1}$, an (asymptotically exact)
mapping onto the familiar action of the Luttinger liquid is
possible. Finally, for weak interactions, a low-order expansion of the
tracelog in $\varphi$ and in the generators of $Q$-fluctuations becomes
permissible. 

Focusing on the latter regime, we apply the formalism to study the
crossover of the tunneling DoS (previously described in
Ref.~\cite{mishchenko}) and the conductivity from the ballistic 
to the diffusive regime.  In the ballistic limit, we recover the 
standard Luttinger liquid power law behaviors in these quantities,
while in the diffusive limit, we obtain a pseudogap in the 
low-energy part of the TDoS and the  $T^{-1/2}$ 
Altshuler-Aronov interaction correction.  
Our main results are Eq.~\eqref{crossdos}
for the tunneling density of states, and Eq.~\eqref{intcorr2}
for the temperature dependence of the conductivity. 
We believe that these results, covering the entire crossover from
ballistic to diffusive, will be valuable  in 
understanding experimental data on nanotubes or nanowires.

Let us conclude by noting some of the open problems in this area.  The
question of what happens to the conductivity at very low $T$ cannot be
answered by our lowest-order calculation.  Relatedly, in order to
treat the small-$N$ limit, new conceptual advances will be necessary.
(Progress along this line has been reported in Ref.\cite{gornyi}.)
Other interesting open questions not addressed
here include the weak localization correction and the
magnetoconductivity, the inclusion of the spin degree of freedom, and
a microscopic calculation of the dephasing time in 1D.  These topics
may be addressed by future work.

\acknowledgments

A.A. thanks J.S. Meyer and A.V. Andreev for discussions.
This work was supported by the DFG-SFB Transregio 12
and by the ESF program INSTANS.

\appendix
\section{Derivation of the effective model Hamiltonian}
\label{1dmic}

In this appendix we review how the model Hamiltonian \eqref{H1d} may
be distilled from  more microscopic descriptions of a quasi-1D
conductor. Generally, the simplifying assumptions below are expected
to hold for energies $\epsilon\lesssim \epsilon_\perp$, where
$\epsilon_\perp$ is the characteristic energy scale related to
transversal (to the cross section of the wire) excitations in the
system. For $l< L_\perp$,
we have  $\epsilon_\perp\sim v_F /L_\perp$, where
$L_\perp$ is a typical transverse width of the wire. 
For $l>L_\perp$, on the other hand, $\epsilon_\perp\sim v_F l/ L_\perp^2$.
 Resolving phenomena on
larger energy scales is a doable task which, however, requires a more
refined (and less universal) modeling. Throughout, the phrase 'low
energy' refers to the regime $\epsilon<\epsilon_\perp$. 

In the absence of disorder and interactions, a quasi-1D conductor is
characterized by $N$ open channels energetically below the Fermi
energy.  The number $N\approx S_\perp/\lambda_F^2$ is often tunable,
e.g., via doping or backgate voltages, where $S_\perp=L_\perp^2$.
We assume that (i) $N\gg 1$ is sufficiently large to justify the
approximations employed later on, 
and (ii) the Fermi energy is located sufficiently far away from the
bottom of the bands formed by longitudinal momentum components to
justify introduction of well-defined chiral (right- or left-moving,
$C=R/L=\pm$) electron branches for each channel; clearly, this becomes
problematic if the Fermi energy is close to the bottom of a
band. Moreover, we shall (iii) focus on the case of spinless (or
spin-polarized) electrons.  The generalization to the spinful case
does not pose conceptual problems and is left to future work.

Denoting the 1D coordinate as $x$ (where we assume that no confinement
along this axis is present) and the transverse degrees of freedom as
${\bf r}_\perp$, the electron operator $\Psi_e({\bf r})$, with ${\bf
  r}=(x,{\bf r}_\perp)$, can be expanded into the $N$ transverse
eigenfunctions as
\begin{equation} \label{expand}
\Psi_e({\bf r}) = \sum_{n=1}^N  \sum_{C=R/L} e^{i C k_{n} x}
 \phi_n({\bf r}_\perp) \psi_{nC} (x) ,
\end{equation}
where $\phi_n ({\bf r}_\perp)$ are the transverse eigenmodes normalized
according to 
\begin{equation}\label{normal}
\int d{\bf r}_\perp \phi^\ast_n({\bf r}_\perp) 
\phi_m({\bf r}_\perp)=\delta_{nm}.
\end{equation}
To give an example, for MWNTs, the eigenmodes, arising from the
wrapping of a 2D graphene sheet onto a cylinder, are \cite{egger}
\begin{equation}\label{mwntconf}
\phi_n(y) = L_\perp^{-1/2} \exp(2\pi i n y/L_\perp),
\end{equation}
where $L_\perp=2\pi R_0$ for outermost-shell radius $R_0$ of the MWNT.
Here, the transverse coordinate is the angular variable $y$, with
$0<y\le L_\perp$.  For given confinement, each band intersects the
Fermi surface at $k=\pm k_n$ with its own Fermi momentum $k_{n}$,
where the slope is given by the respective Fermi velocity $v_{n}$.  At
low energy scales, the kinetic part of the Hamiltonian then is
\begin{equation}\label{h0}
H_0 = -i \sum_C C \int d x \Psi_C^\dagger (x) \hat{v} \partial_x \Psi_C (x),
\end{equation}
with $\hat{v} = {\rm diag} (v_{1},v_{2},
\ldots,v_{N})$.  The noninteracting DoS is thus given by
$\nu_1=\sum_{n=1}^N 1/(\pi v_n)$.
To simplify matters, we neglect differences in the channel-dependent
velocity, $v_{n}\to \langle v_n \rangle_n\equiv v_F$, where $v_F$
denotes the average channel velocity.  As may
be checked, e.g., by an explicit calculation of the diffusive
two-point correlation function of the system, this assumption is
permissible at low energies. For equal Fermi velocities,
the kinetic part of the Hamiltonian assumes the form of $H_0$ in
Eq.~\eqref{H1d}.

Next let us turn to the repulsive Coulomb interactions among the
electrons in the QW.  Starting from an arbitrary microscopic
interaction potential $U_{micr}({\bf r},{\bf r}^\prime)$, which
incorporates external screening effects due to the substrate or
surrounding gates, one arrives at a rather complicated 1D interaction
Hamiltonian.  At low energies, it is however sufficient to
consider a simple model interaction, which is assumed to be
long-ranged (e.g., a $1/r$ potential) on length scales larger than
$L_\perp$.  In that case, the long-range tail of the interaction is
expected to dominate all relevant 1D Coulomb interaction matrix
elements, such as
\begin{eqnarray*} 
U_{nm}(x-x^\prime) &=& \int d{\bf r}_\perp   d{\bf r}_\perp^\prime
U_{micr}(x-x^\prime, {\bf r}_\perp,{\bf r}_\perp^\prime)  
\\ \nonumber &\times&
|\phi_n({\bf r}_\perp)|^2  
|\phi_m ({\bf r}_\perp^\prime)|^2.  
\end{eqnarray*}
For $|x-x^\prime|\gg L_\perp$, the potential $U_{micr}$ is basically
independent of the transverse coordinates ${\bf r}_\perp$ and ${\bf
  r}^\prime_\perp$, implying that the projected interaction does not
depend on the channel indices, $U_{mn}\to U(x-x')$.  The expansion
(\ref{expand}) then leads to an effective contribution to the
Hamiltonian,
\begin{eqnarray}\label{1dinter}
H_{I} & = & \frac{1}{2} \sum_{n,m=1}^N \sum_{C_1,C_2,C_3,C_4}
\int d x d x' U(x-x^\prime) 
\\ \nonumber &\times&
e^{ -i(C_1-C_4) k_n x  } e^{-i(C_2-C_3) k_m x^\prime}
\\ \nonumber &\times&
\psi^\dagger_{nC_1} (x) 
\psi^\dagger_{mC_2}(x') \psi_{mC_3} (x')
\psi_{n C_4} (x).
\end{eqnarray}
Following the standard reasoning leading to LL models
\cite{gogolin,voit}, for a long-ranged potential, electron-electron
(e-e) backscattering ($C_1=-C_4, C_2=-C_3$) processes will be strongly
suppressed. Moreover, e-e Umklapp scattering is important only close
to commensurabilities and will not be discussed here.  For a
long-ranged interaction, the dominant interaction process then
corresponds to e-e forward scattering, where $C_1=C_4$ and $C_2=C_3$
in Eq.~\eqref{1dinter}.  Such processes describe interactions coupling
the 1D density fluctuations,
$H_I = \frac12\int dx dx^\prime \rho(x) U(x-x^\prime) \rho(x^\prime)$.
It is then justified  to express $H_I$ in terms of
the $q=0$ Fourier component of $U(x-x')$, denoted by $U$.  (In
the clean case, up to multiplicative logarithmic corrections in most
observables, this procedure also describes the case of the unscreened
$1/r$ interaction \cite{gogolin}.)  We are then left with the simple
1D forward-scattering interaction \eqref{1dints}. 
We mention in passing that e-e backscattering can be (partially)
included by allowing for different interaction strengths $g_4$ and
$g_2$ in the $\rho_C \rho_C$ and $\rho_C \rho_{-C}$ couplings,
respectively \cite{gogolin}. With minor modifications, our approach
can be adjusted to this situation.  

Finally, quenched disorder is included by starting from a short-ranged
(3D) Gaussian random potential $V({\bf r})$, with the disorder average
defined by its only nonvanishing cumulant
\begin{equation}\label{disav}
\langle   V({\bf r}) V({\bf r}^\prime) \rangle_{dis} =
\frac{1}{2 \pi \nu_3 \tau_0} \delta({\bf r}-{\bf r}'),
\end{equation}
where $\nu_3=\nu_1/S_\perp$ is the 3D noninteracting DoS and
we assume that disorder is weak, $k_n l\gg 1$ for all $k_n$.  With
Eq. (\ref{expand}), we then obtain FS and BS
scattering as in Eqs.~\eqref{hfs} and \eqref{bsd}, respectively,
where 
\begin{eqnarray*}
 \hat{V}_{C,n n'} & = & e^{-i C (k_n - k_{n'}) x } \int d {\bf r}_\perp 
\phi_n^*({\bf r}_\perp)
\phi_{n'} ({\bf r}_\perp) V({\bf r}),
\\
 \hat{W}_{n n'}  &=& e^{-i (k_n + k_{n'}) x } \int d {\bf r}_\perp 
\phi_n^*({\bf r}_\perp)
\phi_{n'} ({\bf r}_\perp) V({\bf r})
\end{eqnarray*}
can be taken as independent random variables whose   statistical
properties follow from Eq.~\eqref{disav}. 
To simplify matters, we assume the specific
torus-type confinement (\ref{mwntconf}). 
Unless one is interested in details of transverse fluctuations or
other high-energy features, the precise form of the confinement
potential is not expected to change the essential physics.
Statistical correlations of $W_{nm}(x)$ can then be described
by the simple form \eqref{dis_cor}. 

\section{Noninteracting limit of the NL${\sigma}$M}
\label{appa}

Let us here  discuss the diffusive properties at $|q|l \ll 1,
|\omega_m|\tau_0\ll 1$ of
the NL$\sigma$M (\ref{nlsm}) in the noninteracting limit.
Using the parametrization (\ref{param}),   
we have two different types of quantum fluctuations, $B_0$ and
$B_1$, where the latter break chiral symmetry. The $B_1$ modes
have a gap, but nevertheless are not to be discarded
since they couple to the massless $B_0$ fluctuations.
We show this now on the Gaussian level, starting from 
the noninteracting version of Eq.~\eqref{nlsm},
\begin{eqnarray} \nonumber
S&=& \frac{\pi\nu_1}{8\tau_0} {\rm Tr} (Q_R Q_L) 
+ \frac{\pi\nu_1}{4} \sum_C {\rm Tr}\Bigl[ 
 D(\partial_x Q_C)^2 \\ 
\label{actnon1}
&& + 2C \Lambda (T_C^{-1}\partial_x T_C) -2 \hat\epsilon Q_C \Bigr],
\end{eqnarray}
where we put $v_F=1$ in intermediate steps.
In order to recover the conventional NL$\sigma$M, we define the
chirally symmetric field $Q_0= T_0 \Lambda T_0^{-1}$ with
$T_0=\exp(-W_0/2)$, and integrate over the massive
modes $B_1$.  Expanding $S$ to second order in $B_1$ (but zeroth order
in $B_0$), the first term in Eq.~\eqref{actnon1} produces
\[
-\frac{\pi \nu_1}{4\tau_0} {\rm Tr}(W_1^2)=
\frac{\pi \nu_1}{2\tau_0} {\rm Tr}(B_1^\dagger B_1^{}),
\]
see Eq.~\eqref{param2}, which
determines the mass gap of the fluctuation mode $B_1$
referred to in Sec.~\ref{seciib}.
The remaining parts of the action \eqref{actnon1} are then expanded to 
linear order in $B_1$, where only the $T^{-1}_C \partial_x T_C$ 
term gives a contribution.  Here one has to be careful because of the
non-commuting nature of $B_{0,1}$, but within a first-order
expansion, no difficulties arise under the parametrization \eqref{param}. 
Terms that are quadratic in $B_1$ or involve spatial derivatives of $B_1$
are neglected since they vanish at low energy and/or
long length scales.  We finally arrive at 
\begin{eqnarray*}
S&=& \frac{ \pi\nu_1}{2} {\rm Tr}\Bigl(
D (\partial_x Q_0)^2 - 2\hat\epsilon Q_0 -W_1^2/2\tau_0 \\
 &-&  iW_1 [ T_0^{-1}\partial_x T^{}_0,\Lambda]_- \Bigr).
\end{eqnarray*}
Next $B_1$ is integrated out, 
and we finally obtain the 1D diffusive action for $Q_0$,
\begin{equation} \label{diffaction}
S =\frac{ \pi\nu_1}{4}\left( D{\rm Tr}(\partial_x Q_0)^2
-4 {\rm Tr}(\hat\epsilon Q_0) \right).
\end{equation}
On the Gaussian level, keeping replica indices implicit, this results in 
\begin{eqnarray*}
S&=& \frac{\pi\nu_1 T^2}{2} \sum_{\epsilon_n>0,\epsilon_m<0}
\int \frac{dq}{2\pi}\\ &\times& \left[ Dq^2+(\epsilon_n-\epsilon_m)\right] 
|B_{0;nm}(q)|^2,
\end{eqnarray*}
implying the usual diffusion pole. 

We finally remark that at energies $\epsilon\agt \epsilon_\perp$, the
diffusion properties (e.g. of MWNTs) become anisotropic. These 'medium
energy effects' may be resolved by employing the full disorder correlator
of the $W_{nm}(x)$,
and decoupling the impurity four-fermion action in
terms of a channel-dependent $Q$-field. One thus obtains a more
structured theory wherein details of anisotropic transport are resolved.

\section{Interaction correction}
\label{appb}

Here we provide the derivation of $S_4^{(3)}$ and
of the corresponding interaction correction, 
see Sec.~\ref{secivb}.
The expansion of Eq.~\eqref{startaction} in $B$ yields
\begin{eqnarray*}
S_B &=& \frac{\pi \nu_1}{8 \tau_0}  {\rm Tr} [ W_1 \Pi_3 W_1 + W_0 \Pi_1
W_0 \\ \nonumber && - W_0 \Pi_2 W_1 + 2 W_0 P_0 + 2 W_1 P_1 ],
\end{eqnarray*}
with kernels 
\begin{eqnarray*}
\Pi_1 (q, \omega) &=&  \frac{v_F^2 q^2 + |\omega| \left( | \omega| + 1/2\tau_0\right) }
{v_F^2 q^2 + (|\omega| + 1 /2 \tau_0)^2}, \\
\Pi_2 (q, \omega) &=&  \frac{q v_F {\rm sgn} (\omega) /\tau_0 }
{v_F^2 q^2 + (|\omega| + 1 /2 \tau_0)^2}, \\
\Pi_3 (q, \omega) &=&  \frac{  \left( |\omega| + 1/\tau_0 \right) 
\left( | \omega| + 1/2\tau_0\right) + v_F^2  q^2 }
{v_F^2  q^2 + (|\omega| + 1 /2 \tau_0)^2}. 
\end{eqnarray*}
Moreover, 
\begin{eqnarray*}
&&
P_0 (q_0,\epsilon_n,\epsilon_n-\omega_0)  = \sum_{C,\omega_m}\int \frac{dq}{2\pi}
\Bigl( \Gamma_C (q,\omega_m)\\ &\times&\Gamma_C(q_0-q,\omega_0-\omega_m) 
 I_C(q,\omega_m,q_0,\omega_0)\\ & +& K_2(q,\omega_m) K_2(q_0-q,
\omega_0-\omega_m) {\rm sgn}(\epsilon_n-\omega_m) \Bigr),
\end{eqnarray*}
and
\begin{eqnarray*}
&&
P_1 (q_0,\epsilon_n,\epsilon_n-\omega_0)  =  \sum_{C,\omega_m} (-iC)
\int \frac{dq}{2\pi}
\Bigl( \Gamma_C (q,\omega_m) \\ &\times& \Gamma_C(q_0-q,\omega_0-\omega_m)  I_C(q,\omega_m,q_0,\omega_0)  \Bigr ),
\end{eqnarray*}
with the kernel
\begin{eqnarray*}
&& I_C(q,\omega_m,q_0,\omega_0)= i v_F\, \int \frac{dp}{2\pi}
G_C(p,\epsilon) \\&& \times G_C(p-q_0,\epsilon-\omega_0)
G_C(p-q,\epsilon-\omega_m).
\end{eqnarray*}
The Gaussian $B$-fluctuations are
then integrated out.  Calling the resulting action $S_4^{(3)}$, we find
\begin{eqnarray*}
&& S_4^{(3)}  = - \frac{\pi \nu_1}{4 \tau_0} 
\sum_{q_0,q_1,q_2}
\sum_{\omega_0>0,\omega_1,\omega_2}\sum_{0\le\epsilon_n\le\omega_0}
\\ &&
\Bigg\{ \Bigg[ \sum_C \Gamma_C(q_1,\omega_1) 
 \Gamma_C(q_0-q_1,\omega_0-\omega_1) \\ && \times
I_C(q_0,\omega_0,q_1,\omega_1) 
\left( 1 - \frac{i C}{4} \frac{\Pi_2(q_0,\omega_0)}{\Pi_3(q_0,\omega_0)} 
\right) \\ &
&  + K_2(q_1,\omega_1) K_2(q_0-q_1,\omega_0-\omega_1) {\rm sgn}(\epsilon_n 
- \omega_1) \Bigg] \\ &
& \times \Bigg[
\sum_{C'} \Gamma_{C'} (q_2,\omega_2) \Gamma_{C'} (-q_0-q_2,-\omega_0-\omega_2) 
\\ &
& \times I_{C'} (q_0,\omega_0,q_0+q_2,\omega_0+\omega_2)  \left( 1 - \frac{i C'}{4} 
\frac{\Pi_2(q_0,\omega_0)}{\Pi_3(q_0,\omega_0)} \right)  \\ &
&  +  K_2(q_2,\omega_2) K_2(-q_0-q_2,-\omega_0-\omega_2) {\rm sgn} 
(\epsilon_n - \omega_0 -\omega_2) \Bigg] \\ &
& \times 
\left( \Pi_1(q_0,\omega_0) - \Pi^2_2(q_0,\omega_0)/4 \Pi_3(q_0,\omega_0) 
\right)^{-1} \\ &
& +   \Bigg[ \sum_C (-i C) \Gamma_C(q_1,\omega_1) 
\Gamma_C(q_0-q_1,\omega_0-\omega_1) \\  &
& \times I_C(q_0,\omega_0,q_1,\omega_1) \Bigg] 
 \Bigg[ \sum_{C'} (-i C') \Gamma_{C'}(q_2,\omega_2) 
\\&& \times \Gamma_{C'}(q_0-q_2,\omega_0-\omega_2) \\ &
& \times  I_{C'}(q_0,\omega_0,q_0+q_2,\omega_0+\omega_2) 
\Bigg] / \Pi_3(q_0,\omega_0) \Bigg \}.
\end{eqnarray*}
Equation \eqref{corr1} is then used to compute
the conductivity
correction. The contraction of $\varphi(q,\omega)\varphi(-q,-\omega)$ 
with $S_4^{(3)}$ 
leads to eight terms, with four
different terms appearing twice each. These four terms are in fact two by
two equivalent via the symmetry $\omega \to -\omega$.
In particular, the two contributions with $-\omega$ vanish after the
summation over $\epsilon_n$ is performed.
The two possible contractions are (i) $q_1 = -q_2 = q$, $\omega_1 = -\omega_2
= \omega$, and (ii) $q_1=q$, $\omega_1 = \omega$, $q_2 = q - q_0$, $\omega_2
= \omega - \omega_0$. 
The external momentum $q$ is now taken to zero, and we also let
$\omega\to 0$ under the constraint ${\rm Re}\,\omega >0$.
The $\epsilon_n$-summation can be done easily
and yields, after the frequency shift  $\omega_0 \to \omega_0 + \omega_m$,
\begin{eqnarray*} 
\delta  \sigma_3 & =& -\frac{8 v_F^2 e^2 \nu_1 \tau_0^3}{\omega_m} \sum_{\omega_0>0,q_0}
 \frac{ \omega_0+\omega_m - 
| \omega_0-\omega_m | }{\Pi_0(q_0,\omega_0)} \\
& \times& \bigg[ X_1^2 (q_0,\omega_0) \left( \Pi_1(q_0,\omega_0) 
- \Pi^2_2(q_0,\omega_0)/4 \Pi_3(q_0,\omega_0) \right)^{-1}\\
& + & X_2^2(q_0,\omega_0) / \Pi_3(q_0,\omega_0) \bigg],
\end{eqnarray*}
where $\omega_m \to 0$ has been anticipated and 
we have used the abbreviations
\begin{eqnarray*}
& & X_1 (q_0,\omega_0)  = \frac{1}{4 \tau_0^2} \sum_C \frac{C \Gamma_C(q_0,\omega_0)}{2 i v_F q_0 \varphi
(q_0,\omega_0)} \\
&& \times  \left( 1 - \frac{i C}{4} \frac{\Pi_2(q_0,\omega_0)}{\Pi_3(q_0,\omega_0)} \right) 
I_C^0(q_0,\omega_0) + \frac{i}{2 \tau_0} 
\frac{K_2 (q_0,\omega_0)}{i v_F q_0 \varphi
(q_0,\omega_0) },  \\ &
& X_2 (q_0,\omega_0)  = \frac{1}{4 \tau_0^2} \sum_C \frac{(-i) \Gamma_C (q_0,\omega_0)}{ 2 i v_F q_0   \varphi
(q_0,\omega_0)} I_C^0(q_0,\omega_0), \\ &
 & I_C^0(q_0,\omega_0)  = 2 \tau_0 \, \frac{1/\tau_0 + \omega_0 - i C v_F q_0}
{(1/2\tau_0 + \omega_0 - i C v_F q_0)^2}.
\end{eqnarray*}
The summation over $\omega_0$ is then replaced by a contour integral,
with a line cut for $\Omega =  i \omega_0 = i\omega$. Performing the analytic
continuation $i \omega = \omega' + i 0^+$
with $\omega' \to 0$,
we finally obtain $K_3$ in Eq.~\eqref{conduct_total}.


\begin{thebibliography}{99}
\bibitem{footint} Once
interchain hopping becomes relevant, coherent electron transfer among
the chains is important and eventually induces a crossover into a
Fermi liquid state.  Such effects are implicitly contained in the
model studied below, since the $N$ 'open channels' directly represent
the eigenmodes of the noninteracting clean system.
\bibitem{foot_boson} The reason for avoiding bosonization is that a
  straightforward multi-channel generalization of the bosonized
  theory~\cite{giamarchi} does not properly describe the diffusive
  phase in the noninteracting limit. Technically, this failure can be
  traced to a breaking of the replica rotation symmetry originally
  present in the non-interacting model, yet absent in the (abelian)
  bosonized theory~\cite{giamarchi}. For related comments, see also
  Ref.~\cite{gornyi}.
\bibitem{footgauge} For channel-dependent interactions beyond
  Eq.~(\ref{1dints}), and for non-uniformly distributed Fermi
  velocities additional terms may be generated. However, we do not
  believe these terms to be of physical relevance. Specifically, on
  the noninteracting level, and independently of fluctuations in the
  Fermi velocities, the correct (anisotropic) diffusion properties are
  recovered from impurity backscattering alone.
\bibitem{footgreen}
In the diagrammatic 
approach, the disordered Green function has a $1/2\tau_0$ prefactor  
before $\Lambda$, in contrast to Eq.~\eqref{disordergr}. 
The difference arises here quite naturally.
\bibitem{foot1}
There is an additional prefactor 
$K^{1/N}$ in Eq.~\eqref{crossdos}, which we
drop here.  For $N\gg 1$ and $\nu_1 U\ll 1$,
it is very close to unity.
\bibitem{footgia}
The critical value $K=3/2$  follows
by taking into account the disorder-induced
renormalization of the interaction.
This effect is only important for very low temperatures,
a regime not considered in this paper.
\end{thebibliography}
\end{document}